# DYNAMO DRIVEN ACCRETION DISCS AND DWARF NOVA ERUPTIONS

P.J. Armitage[1]

M. Livio[2]

J.E. Pringle[1,2]

## ABSTRACT

We explore the consequences of a magnetic dynamo origin for the viscosity in accretion discs, for the structure and evolution of discs in dwarf nova systems. We propose that the rapid cooling that sets in at the end of a dwarf nova eruption acts to inhibit the Balbus-Hawley instability, and thereby to quench dynamo action and so reduce disc viscosity. We demonstrate that a modified disc instability model can reproduce the basic properties of dwarf nova eruptions, as well as some properties of quiescent discs. We also discuss some observational consequences of our model.

## 1. INTRODUCTION

The dwarf novae, a subset of the cataclysmic variables, offer the best conditions in which to study the uncertain physics of accretion discs. This is a consequence both of their origin in binary systems (allowing determination of many of the important parameters of the system, and, in fortuitous examples, observation of eclipses), and the fact that most of the disc luminosity is radiated at accessible wavelengths. Equally importantly, the repeated outbursts seen in these systems, which typically last a few days, permit the exploration of the disc's time-dependent behaviour.

The disc instability model is able to account for many, but not all, of the properties of dwarf nova outbursts. In particular, the X-ray fluxes of the quiescent disc and the observed delay of the UV flux relative to the optical on the rise to outburst remain problematic (Livio 1994). A more fundamental worry is that such models require that the Shakura-Sunyaev $\alpha$ viscosity parameter be different in outburst and in quiescence, typically by a factor of three or more. Although there is no reason why $\alpha$ should *not* take this (or, indeed, any other) functional form, the *ad hoc* nature of the requirement illustrates clearly the limitations imposed by our ignorance of the mechanism that gives rise to the disc viscosity.

The possibility that small-scale magnetic fields might provide a mechanism for angular momentum transfer in discs was recognized by Shakura & Sunyaev (1973). More recently, mechanisms that might generate and sustain such fields by the action of a magnetic dynamo in the disc have been proposed by various authors (e.g. Balbus & Hawley 1991,1992; Tout & Pringle 1992; Brandenburg et al. 1995). In the Tout & Pringle scenario, dynamo action occurs as a consequence of shear, reconnection, and the Balbus-Hawley and Parker instabilities. A merit of such a model is that it relies only on physical processes that are well-established, although much further study is required to clarify their detailed combined operation in the context of accretion discs. This, together with the fact that the model does not require any pre-existing turbulence in the disc for its operation, suggests that such a dynamo is a promising candidate for the origin of the elusive viscosity.

In this paper we explore the consequences of a magnetic dynamo origin for the viscosity in dwarf nova discs. Using a time-dependent disc code, we demonstrate that a modified disc instability model can reproduce the basic properties of dwarf nova outbursts. In our model, rapid cooling at the end of an outburst acts to inhibit the Balbus-Hawley instability and quenches dynamo action. The disc remains in a quiescent state, with strong fields (relative to the thermal energy) and very low viscosity until sufficient mass has been added to restart the dynamo at the outer disc edge. We also discuss the observational consequences of our model.

The plan of this paper is as follows. Section 2 outlines the aspects of the dynamo model that are relevant for this work, and how periodic outbursts can be generated within the model. Section 3 describes the time-dependent disc code, and in Section 4 we present model outburst calculations. Section 5 discusses further observational consequences of the model.

## 2. THE DYNAMO MODEL

The ingredients required to sustain a magnetic dynamo in a disc are differential rotation (to generate azimuthal field from radial), and some mechanism that regenerates radial field from azimuthal, thereby closing the feedback loop. The first step is straightforward—in a disc rotating with Keplerian angular velocity, shear creates azimuthal field from radial on a dynamical timescale—but the second is more problematic. Stellar type dynamos that rely on convective motions to close the cycle are unattractive, as many astrophysical discs are expected to be stable against convection. Mechanisms that rely on pre-existing turbulent disc flows are similarly unappealing, as no purely hydrodynamic instability has been shown to exist in accretion discs.

In the Tout & Pringle (1992) dynamo cycle, the crucial step of regenerating radial field is accomplished by the Balbus-Hawley instability. This is a linear, local instability that exists in rotating flows threaded by an arbitrarily weak vertical field, for which $d\Omega^2/dR < 0$, conditions satisfied in discs (Balbus & Hawley 1991, 1992; Hawley & Balbus 1991, 1992). The dynamo cycle then involves shear (to create azimuthal field from radial); the Parker instability (to turn azimuthal field into vertical); and the Balbus-Hawley instability (to create radial field from vertical). Reconnection of opposite sign patches of vertical field near the disc surface provides a dissipative mechanism, and limits the growth of $B_Z$.

Numerical simulations show, unsurprisingly, that the magnetic fields generated by the interplay of these processes are extremely complex, both spatially and temporally (Hawley, Gammie & Balbus 1995; Stone & Norman 1994; Brandenburg et al. 1995). However, working in terms of local quantities $B_R$, $B_\phi$ and $B_Z$ (which one hopes can be regarded as appropriate spatial averages of the actual fields), it is possible to make estimates of the rates of the various instabilities and so to investigate semi-quantitatively the dynamo action. This model is undoubtably oversimplified. However, the results do imply that a magnetic dynamo of this type could produce a Shakura-Sunyaev 'alpha' parameter in the 0.1–1 range typically inferred to exist in dwarf nova disc (Tout & Pringle 1992). The fields, and hence also the equivalent Shakura-Sunyaev 'alpha' parameter, $\alpha_{SS} = (B_R B_\phi / 4\pi\rho C_s^2)$, where $\rho$ is the density and $C_s$ the sound speed, exhibit periodic oscillations. The timescale for these is $\sim 10\Omega^{-1}$, i.e. an order of magnitude slower than the dynamical timescale of the disc and comparable to the thermal timescale $t_{\rm th} \sim \alpha^{-1}\Omega^{-1}$. We therefore propose a model in which the cooling at the end of an outburst may be sufficiently rapid that the dynamo is thrown out of equilibrium. The sound speed and disc scale height drop rapidly, leaving behind a strong vertical field that is unable to decay as promptly. The growth rate of the Balbus-Hawley instability is *zero* if,

$$B_Z^2 > \frac{24}{\pi}\rho C_s^2. \tag{1}$$

Thus, rapid cooling can shut off the instability, and the turbulent motions generated by it, entirely.

Although the vertical field exterior to the disc may now pinch off and escape, the reconnection and decay of the field actually threading the disc material is liable to be slow, because there is no longer dynamo generated turbulence to stir up the disc and push field regions of opposite sign together. We therefore propose that the disc may enter into a prolonged quiescent state, with strong but stable vertical fields and very low viscosity. This phase is ended only when sufficient mass (and energy) has been added at the outer edge to violate inequality (1) and allow the Balbus-Hawley instability to restart. There is then the potential for a global limit cycle in which the disc switches between active (high viscosity), and quiescent dynamo states. If the active dynamo is able to generate an effective $\alpha \sim 0.1\text{-}0.3$ (and here we assume that this is the case), then such a limit cycle leads to outbursts of the type seen in dwarf nova systems.

We note at this stage that the structure of the quiescent, strongly magnetised disc is highly uncertain. A disc in which the Balbus-Hawley instability is inhibited will cool during inter-outburst to a state where the thermal contribution to the energy and pressure is small compared to the magnetic contribution. It has been suggested that such a disc might break-up into separate magnetised blobs (Pringle 1981) whose collisions would be cushioned by the strong fields. Such a disc, like the continuous fluid disc envisaged above, would have a low apparent $\alpha$, but in other observable aspects (eg cooling time) would differ considerably. Speculating further, one can imagine that during the long period of quiescence, the dynamo fields, which when generated will be small scale (of order the disc scale height), might reconnect in the disc corona to produce a more ordered global field (Tout & Pringle 1995). The disc could then drive a magnetic wind that would both carry material out of the disc plane and remove angular momentum from the inner disc. Accretion of 'cold' disc material could then occur even in the absence of any significant viscosity, leaving an almost evacuated hole in the inner disc. It is not clear how to model these processes quantitatively, and so for our models we regard the quiescent disc as continuous and devoid of a wind. However, as we discuss in section 5, more radical ideas of this type do show promise in explaining some of the puzzles thrown up by observations of dwarf novae.

## 3.  THE DISC MODEL

To follow the global evolution of the disc, we use a one-dimensional vertically averaged disc code. The evolution of the surface density $\Sigma$ is described by the combined equation for mass and angular momentum conservation,

$$\frac{\partial \Sigma}{\partial t} = \frac{3}{R} \frac{\partial}{\partial R} \left[ R^{1/2} \frac{\partial}{\partial R} \left( \nu \Sigma R^{1/2} \right) \right], \tag{2}$$

where the kinematic viscosity $\nu = \alpha C_S^2 / \Omega$. We follow the disc central temperature $T$ via an energy equation in the form,

$$\frac{\partial T}{\partial t} = \frac{2(Q_+ - Q_-)}{C_V \Sigma} - v_R \frac{\partial T}{\partial R} + (\Gamma_3 - 1) \frac{T}{\Sigma} \frac{\partial \Sigma}{\partial t} + (\Gamma_3 - 1) \frac{T}{\Sigma} v_R \frac{\partial \Sigma}{\partial R}. \tag{3}$$

Here $C_V$ is the specific heat capacity at constant volume and $\Gamma_3$ is the third adiabatic exponent. In the first term on the right hand side, $Q_+$ and $Q_-$ represent the effects of viscous heating and radiative cooling from the disc surfaces. We have,

$$Q_+ = \beta \nu \Sigma \Omega^2 + \sigma T_{\text{floor}}^4, \tag{4}$$

$$Q_- = \sigma T_e^4, \tag{5}$$

where $T_e$ is the effective temperature of the disc surface and $\sigma$ the Stefan-Boltzmann constant. The parameter $\beta = (9/8)$ if all the energy dissipated goes into local heating of the disc. For simplicity we assume this in our calculations, but note that $\beta$ could take a lower value as a consequence of energy being injected into a disc corona or as kinetic energy in a wind. The term $\sigma T_{\text{floor}}^4$ is included to prevent the disc from ever cooling below some background temperature, and represents crudely external heating of the disc from the white dwarf and secondary star. The second term represents advection of heat due to the radial flow at velocity

$$v_R = -\frac{3}{\Sigma R^{1/2}} \frac{\partial}{\partial R} (\nu \Sigma R^{1/2}), \tag{6}$$

while the remaining two terms account for the energy change from $PdV$ work. We do not include terms for radial radiative diffusion or non-local transport of energy from the turbulence associated with the viscosity, as have been incorporated in some other studies (e.g. Cannizzo 1993; Faulkner, Lin & Papaloizou 1983). The omission of these terms alters the outburst behaviour, but is not likely to change the general characteristics of the outbursts (Cannizzo 1993).

Previous studies (Mineshige 1988) have found that the sharp increase in $C_V$ associated with the ionization of hydrogen (at $T \sim 10^4 K$) makes a significant difference in dwarf nova models, by slowing the transition to the upper state. To include this effect, we employ the analytic expression for the heat capacity at constant pressure, $C_P$, given in Cannizzo (1993),

$$\frac{C_P(\rho, T)}{\mathcal{R}/\mu} = 2.7 + A e^{-(\log T - \log T_0)^2/w^2}, \tag{7}$$

where $\mu$ is the mean molecular weight, $\mathcal{R}$ is the gas constant, $A = 26 - 7(\log \rho + 7)$, $\log T_0 = 4.1 + 0.07(\log \rho + 7)$, and $w = 0.115 + 0.015(\log \rho + 7)$. This gives a good fit over the relevant density range to numerical calculations. The specific heat capacity at constant volume is then given by $C_V = C_P - \mathcal{R}$.

To relate the central temperature $T$ to the effective temperature $T_e$, we utilise the relations given by Cannizzo (1993) for the hot state, which are based on vertically averaged Shakura-Sunyaev type scalings. For a disc around a star of mass $M_1 = M_*/M_\odot$,

$$T_e = 3.5 \times 10^4 M_1^{-1/8} R_{10}^{3/8} \Sigma_2^{-1/2} \mu^{-1/8} T_5^2 \mathrm{K}, \tag{8}$$

where $R_{10} = R/10^{10} \mathrm{cm}$, $\Sigma_2 = \Sigma/100 \mathrm{g cm}^{-2}$, and $T_5 = T/10^5 \mathrm{K}$. A hot state as defined by this power law (equation 8) exists down to a minimum surface density

$$\Sigma_{\min} = 8.25 R_{10}^{1.05} M_1^{-0.35} \alpha^{-0.8} \mathrm{g cm}^{-2}, \tag{9}$$

where $\alpha$ is the value appropriate to the high state of the disc.

When a zone in the disc is *not* in outburst, equation (8) will certainly fail to provide an accurate representation of the $T$-$T_e$ scaling. However, we cannot simply adopt one of the many low state vertical structure calculations, because the quiescent state we envisage is entirely different from that normally assumed, being out of thermal equilibrium and dominated by strong magnetic fields. In the absence of a consistent model for this regime, we choose to use equation (8) throughout the cycle, thereby ignoring the complexities of the low state altogether. This is wrong for two distinct reasons. First, it means that we model the cooling of the quiescent disc inaccurately, and should therefore treat details of our $T_e$ profiles in the low state with caution. The robust prediction–that in the absence of viscosity $T_e$ will rapidly cool below the temperatures seen in standard models–is however preserved by using our prescription. Second, we ignore the possible existence of a low branch that would provide a stable low state when the dynamo is first restarted. This would provide an additional delay to the transition back to the hot state, over and above that provided by the requirement that inequality (1) be violated. Our models do not *require* such an additional delay to produce well separated outbursts. If it were present, however, limit cycle behaviour would persist, albeit with differences in the recurrence time and accretion rate of outbursts from those presented here.

Mass is added to the outer edge of the disc at a rate $\dot{M}_{\mathrm{outer}}$, and with specific angular momentum corresponding to a Keplerian orbit at radius $R_h$. We employ the method used by Cannizzo (1993) and add the mass flux from the secondary in a Gaussian distribution centred at $R_h$, with width $\Delta R_h$. The mass added is assumed to have zero magnetic flux and the same temperature as the material at the outer edge of the disc. We note, however, that in this model any variations in the magnetic flux of the stream material (e.g. due to magnetic cycles on the secondary) could directly influence the triggering of outbursts by altering the flux in the outermost regions of the disc. As has been discussed by other authors (e.g. Meyer-Hofmeister, Vogt & Meyer 1995), the field of the secondary may also play a role in dwarf novae by threading the disc and removing angular momentum.

For the magnetic field evolution we adopt a straightforward approach. We assume that when averaged azimuthally, the dynamo produces a viscosity that can be represented by a constant value of $\alpha$. The vertical magnetic field is then given approximately by the value at which the growth of the Balbus-Hawley instability is curtailed,

$$B_Z^2 = \frac{24}{\pi} \rho C_s^2. \tag{10}$$

When the dynamo is operating in a given zone, the value of $B_Z$ in that zone is set to this value at each timestep according to the local values of the density and sound speed. The dynamo is switched off in a zone when the surface density $\Sigma$ falls below the minimum value for the hot branch $\Sigma_{\min}$ as given by equation (9). When this occurs, the value of $B_Z$ is 'frozen' and $\alpha$ is set to an arbitrary low value (typically $10^{-4}$, which for numerical reasons is preferable to actually taking zero). The dynamo is restarted when the density and sound speed have increased to the level where inequality (1) is violated, and $\alpha$ is reset to its high value.

The numerical method is as described by Pringle, Verbunt & Wade (1986). The disc between $R_{\rm in}$ and $R_{\rm out}$ is divided into $N$ zones evenly spaced in $\sqrt{R}$. The boundary condition at the inner edge is that $\Sigma(R_{\rm in}) = 0$, and at the outer edge that $v_R = 0$. The disc evolution equations (2) and (3) are integrated using an explicit first-order scheme, with the advective terms being treated using Lelevier ('upwind') differencing (Potter 1973). The calculations described here use $N = 100$. Previous authors have argued that this should be sufficient to resolve the transition front adequately (Cannizzo 1993; Lin, Papaloizou & Faulkner 1985).

## 4. MODEL OUTBURST CALCULATION

Given the present uncertainties in the model set out in Sections 2 and 3, it would be premature to attempt to match observations of a specific system. Our aim is rather to demonstrate that with a plausible choice of parameters, outbursts similar to those observed can be generated, and to discuss the general differences from the standard picture expected in a magnetic dynamo driven model. We take as system parameters a white dwarf of radius (equal to the inner disc radius) $R_{\rm in} = 10^9$cm, with a mass $M_* = 0.8 M_\odot$. The outer edge of the disc is at $R_{\rm out} = 3 \times 10^{10}$cm, $R_h = 2.6 \times 10^{10}$cm, and $\Delta R_h = 4 \times 10^9$cm. Mass is added to the disc at a constant rate $\dot{M} = 10^{16}$gs$^{-1}$. The model is run until quasi-steady limit cycle behaviour is obtained.

For the internal model parameters, we take the Shakura-Sunyaev viscosity parameter in the high state to be $\alpha = 0.25$. The high state $\alpha$ controls the duration of outbursts in the usual manner. We take $T_{\text{floor}} = 10^2$K (though in practice the disc remains well above this temperature throughout the cycle), $\mu = 0.615$, and $\Gamma_3 = 5/3$.

## 4.1.   Properties of Model Outbursts

Figures 1 and 2 show the accretion rate onto the white dwarf and the luminosity of the system as a function of time for a calculation with the above parameters. As can be seen from the Figures, the system undergoes outbursts, with a variable recurrence time of 5–10 days. Each individual outburst lasts for 3–5 days, with a rise to maximum of approximately 1 day. The strongest outbursts have a maximum accretion rate of approximately $10^{17}$gs$^{-1}$, with a decay from maximum that is faster than exponential. The accretion rate during the quiescent phase is $\dot{M} \approx 10^{11}$gs$^{-1}$, and is determined entirely by the arbitrary value of $\alpha$ in the low phase. The accretion rate onto the central object can be taken as a measure of the boundary layer luminosity. In contrast we also show the bolometric luminosity of the disc. We have included a contribution from the bright spot where the incoming stream strikes the outer edge of the disc, in order to give an estimate of the magnitude of the outbursts from an observational point of view (cf. Pringle, Verbunt & Wade 1986). For the parameters of this run the bright spot luminosity is estimated at $\sim 10^{31}$ergs$^{-1}$. During interoutburst, the luminosity from passive cooling in the disc is typically at least an order of magnitude below this, so that the quiescent disc luminosity is dominated by the bright spot. The general appearance of these outbursts is very similar to those obtained from standard non-magnetic models (e.g. Cannizzo 1993; Ichikawa & Osaki 1992).

To illustrate the principal difference of this model from the standard ones, *viz.* the lack of a lower branch to the 'S-curve,' we plot in Fig. 3 the path traced out in the $\log(\Sigma/\text{gcm}^{-2})$-$\log(T_e/\text{K})$ plane by a single annulus in the disc, here at a radius $R = 2 \times 10^9$cm. Points are plotted over the course of two outburst cycles. The cycle begins with a cold quiescent disc cooling at constant surface density below 1000K (labelled as 1 in Fig. 3). As the outburst commences dynamo action resumes in the next zone out (2). Mass and energy are then advected into the annulus, increasing $\Sigma$ and heating it up until inequality (1) is violated and dynamo action is restarted (3). The annulus then climbs rapidly to join the hot branch (4), after which it continues to heat up, reaching a peak temperature (5) that depends on the strength of the outburst (here we show two fairly small outbursts). The annulus then remains on the hot branch until $\Sigma$ falls below $\Sigma_{\text{min}}$ (6), at which point the outburst for this zone comes to an end. The dynamo switches off and the annulus resumes cooling, though until the neighbouring zone on the inside also becomes quiescent, mass and energy continue to be advected in. This can cause the dynamo to switch on again briefly until the neighboring zones have also become quiescent, after which the zone we are following just cools at constant $\Sigma$. The value of $\Sigma$ in quiescence is then approximately $\Sigma_{\text{min}}$ at the radius of that zone, though the presence of advection into a zone after the dynamo there has switched off means it can be somewhat higher. This is shown in the Figure, where it can be seen that the second outburst cycle has a higher frozen value of the surface density than the first.

## 4.2. The Disc in Quiescence and in Outburst

Figures 4–6 show the radial variations of the surface density, the vertical magnetic field and the effective temperature during the rise to outburst and during the return to quiescence. Curves which correspond to significant points in the cycle for a zone in the inner disc are labelled to correspond with Fig. 3; (1) the quiescent disc, (2) when the heating wave reaches the inner disc, (5) the peak of the outburst, and (6) when the dynamo is about to switch off in the innermost zones. In quiescence the lack of viscosity means that the surface density profile is strongly weighted to large radii. The fraction of the disc mass accreted during outburst is $\sim 2\%$ for the small outbursts and $\sim 10\%$ for the larger events. Throughout the cycle most of the mass in the disc remains at large radii (Figure 4).

During quiescence the frozen magnetic field $B_Z$ reaches a few $\times 10^3$ G at all radii in the disc. Substantially stronger fields are obtained during outburst, with a maximum value of $\sim 30$ kG being attained near the white dwarf at the peak of the outburst (Figure 5). The fields in quiescence are lower because the disc cools substantially below its peak values before the dynamo ceases operation. We also note that as the surface density in the inner disc remains essentially fixed in quiescence, outbursts are forced to start from the outer edge and propagate in. This is in contrast to the standard model where either inside-out or outside-in outbursts may occur, depending primarily on the mass addition rate (Ichikawa & Osaki 1993).

Figure 6 shows the run of the disc effective temperature as a function of time. During outburst the values of a few $\times 10^4$K are similar to those obtained by other authors (eg Cannizzo 1993) for the inner disc regions. This is not surprising since we have chosen parameters to give us similar accretion rates, principally by using the same upper (hot) branch of the S-curve. During quiescence the temperatures are a factor of 2–3 lower, due to the lack of internal heating. Indeed in quiescence in our models we find that the temperature profile is essentially flat with radius. This appears to be a consequence of two competing effects–the higher surface density at large radii leads to a longer cooling time, but this is compensated by the fact that during outburst the inner regions of the disc are heated to a much higher temperature. This result should be treated with caution however, since we have modelled the cooling properties of the quiescent disc only in an approximate manner (Section 3).

## 5. OBSERVATIONAL CONSEQUENCES OF A DISC DYNAMO

In the previous sections we have outlined a model for dwarf nova outbursts based on a magnetic dynamo origin for the disc viscosity. In our model, the low viscosity at quiescence is a direct consequence of the quenching of dynamo action in a rapidly cooling disc. This, in turn, is due to the fact that as the sound speed drops, while the magnetic field does not decay (or does not do so as rapidly), the disc becomes stable to the Balbus-Hawley instability, which is a crucial ingredient to field generation in the dynamo. The resultant model is able to reproduce the periodic outbursts that are the most distinctive feature of these systems. In this Section we seek to point out further observational consequences of the model, with particular reference to the fact that in this model the transport of angular momentum in the disc is dynamo dominated.

An immediate consequence of the model is that, at least in the absence of a disc wind, which is in principle capable of giving rise to angular momentum loss, the accretion rate during quiescence should be very low and emission from a boundary layer largely absent. Regardless of whether a wind exists or not, the inner parts of the disc are expected to be much cooler than in the standard case, leading to a much reduced UV flux from the central regions of the accretion disc. This deficit of UV emission from the central areas is consistent with HST observations of IP Peg discussed by Horne (1994). We also find in our models (Figure 6) that during quiescence the effective temperature of the disc is essentially constant with radius (although we have warned that the input physics for this part of the calculation is rather approximate). Even so, it is worth noting that observations of the dwarf novae OY Car and Z Cha in quiescence (Wood et al. 1986; 1989), show flat radial temperature profiles. It is also found that in quiescence the mass transfer rate in OY Car appears to be two orders of magnitude smaller in the inner disc than near the disc edge, suggesting a very low viscosity in the quiescent state.

During outburst, the accretion rate and other properties are similar to those predicted by the standard models. However, we might expect to observe flickering in the UV emission as a result of the fluctuating nature of the dynamo process, at least in those systems where the disc reaches the white dwarf surface. To estimate this effect, we note that the typical scale of dynamo generated fields is $\sim 2H$, the disc thickness. At the inner edge of the disc during outburst, the ratio of the disc scale height to the radius ($H/R$) is a few $\times 10^{-2}$, and so there will be $\sim 10 - 10^2$ separate dynamo 'cells' at the inner edge of the disc. In each cell the effective viscosity fluctuates on a timescale that is longer than the local thermal timescale, implying that luminosity changes at the $\leq 10\%$ level, arising from Poisson fluctuations in average dynamo activity and accretion rate, may be observable during outburst.

When the dynamo is operating, the potential energy of the disc material is being dissipated via expulsion of flux loops from the disc, followed by reconnection in the more tenuous atmosphere above the disc surface. The field strengths involved are of order 10kG, and cover the entire disc surface. We might therefore expect to observe chromospheric lines of species such as Ca II and Mg II that are being powered non-radiatively by Alfvén waves and reconnection events, together with an extensive disc corona at temperatures of $10^{6-7}K$ that emits in the X-ray region of the spectrum. Recent ASCA observations of SS Cygni in an anomalous outburst (Nousek et al. 1994) do seem to detect emission that cannot be explained by simple boundary layer models, and this could be due to a coronal X-ray component at intermediate temperatures. Similarly, it should be noted that recent models for AGN discs also suggest that most of the energy dissipation occurs in a corona at low optical depth, and that this corona in turn irradiates the disc (Zycki et al. 1994, and references therein). The general prospects of detecting the kind of anisotropic turbulence one would associate with a dynamo model of this sort, through observing changes in the profiles of double peaked emission lines were discussed recently by Horne (1994).

During quiescence, the magnetic field actually threading the disc matter is frozen. However, the strong fields 'left behind' when the disc scale height collapses at the end of an outburst can still power a lower level of coronal activity, and perhaps drive a wind. For the magnetic field strengths and disc volumes of the model presented earlier, the total magnetic energy available in these fields is approximately $10^{35-36}$ erg. During quiescence this energy may be released in the form of X-ray flares as large loops of flux are formed and interact in the coronal regions. The relevant velocity in such a corona is the Alfvén speed, and so we might expect to see spectral lines broadened to velocities much greater than expected on the basis of the sound speed. We note that the prominent Fe absorption lines seen in OY Car (Horne 1994) show turbulent broadening characteristic of a Mach number of order 10.

The possibility of a magnetic wind in at least some systems is a further attractive speculation. Such a wind would both carry disc matter through our line of sight far from the disc plane, and act to remove angular momentum from the inner disc and deplete the surface density. Most theoretical attempts to reproduce the observed delay of the UV flux relative to the optical on the rise to outburst are based on creating just such a 'hole' in the inner disc, typically either by white dwarf magnetic fields or the action of a 'siphon flow' that clears the inner disc (e.g. Livio & Pringle 1992; Meyer & Meyer-Hofmeister 1994). A magnetic wind that predominantly depleted the innermost disc region of matter would act in a similar, but potentially more dramatic way to these mechanisms.

To conclude, dynamo driven accretion discs appear to offer for the first time a physical model for angular momentum transport. The present work opens up the possibility that dwarf nova eruptions, one of the most dramatic manifestations of disc instabilities, can be incorporated into the context of dynamo driven discs.

## Acknowledgements

ML acknowledges support from NASA grant NAGW-2678. JEP acknowledges support from NATO Travel Grant CRG 940189, and thanks the Space Telescope Science Institute for hospitality.

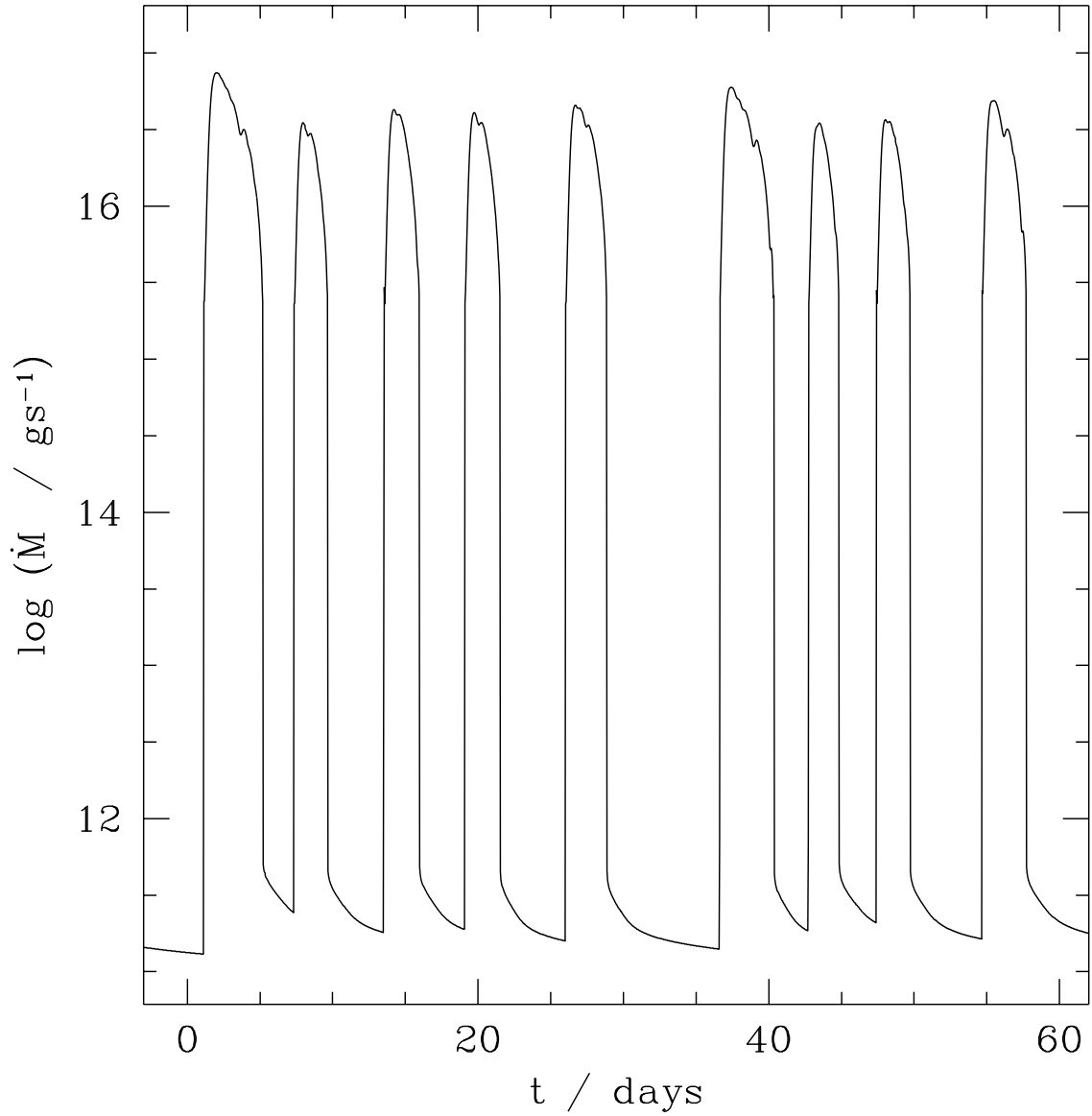

Fig. 1.— Log of the accretion rate (gs⁻¹) through the inner edge of the disc plotted as a function of time (days), obtained after a long run.

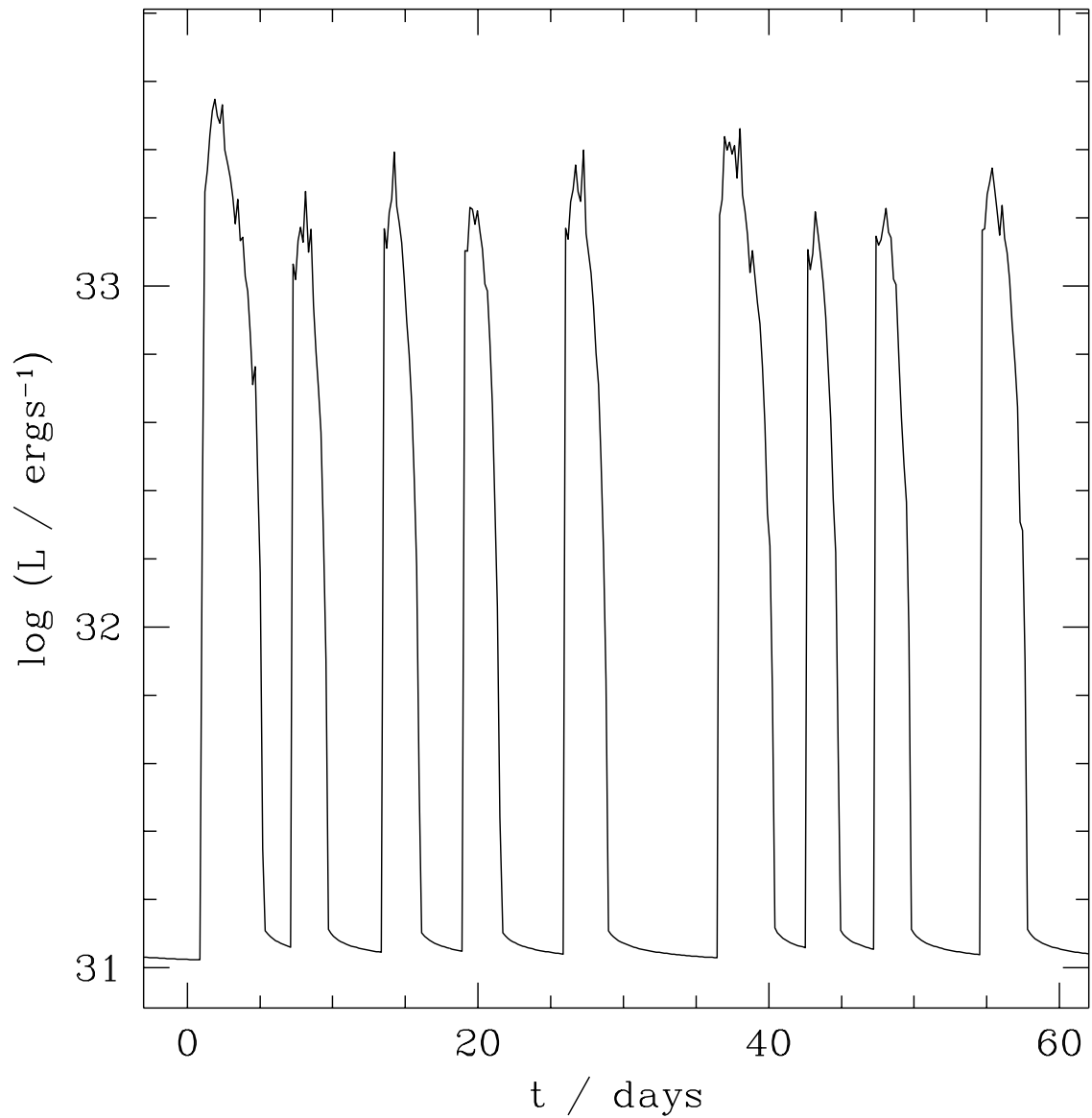

Fig. 2.— Plot of log disc luminosity (ergs⁻¹) as a function of time (days), corresponding to the outbursts shown in Fig. 1. The disc luminosity is calculated from integrating the local luminosity across the disc surface, plus the contribution from the bright spot. The boundary layer luminosity is not included.

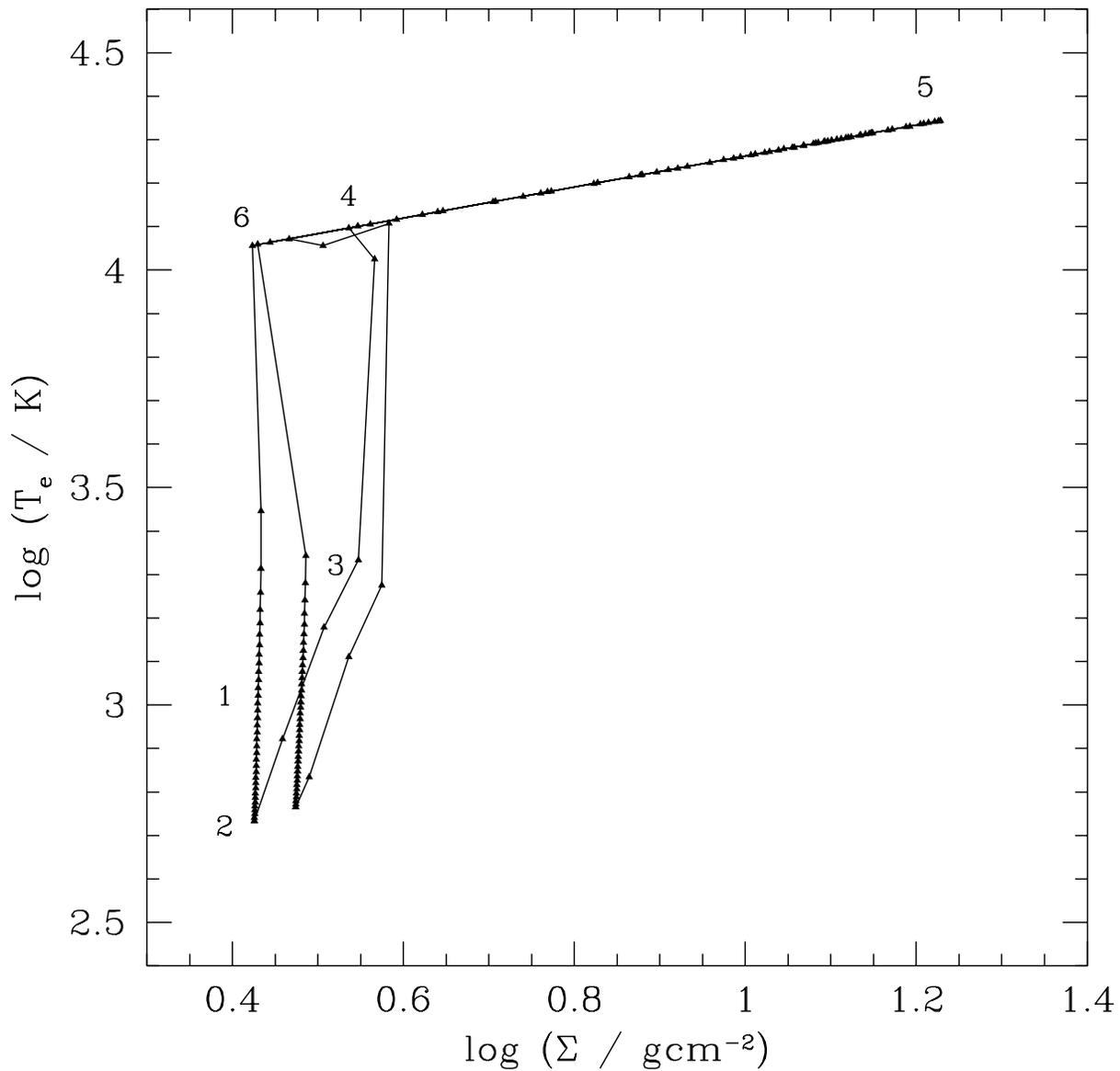

Fig. 3.— Path traced out in the $\log(\Sigma/\mathrm{gcm^{-2}})$ - $\log(T_c/\mathrm{K})$ plane by an annulus at $R = 2 \times 10^9 \mathrm{cm}$. Points are plotted at 50s intervals during the rise to the hot branch, and at $5 \times 10^3$s intervals subsequently. The numeric labels refer to significant points in the cycle described in the text.

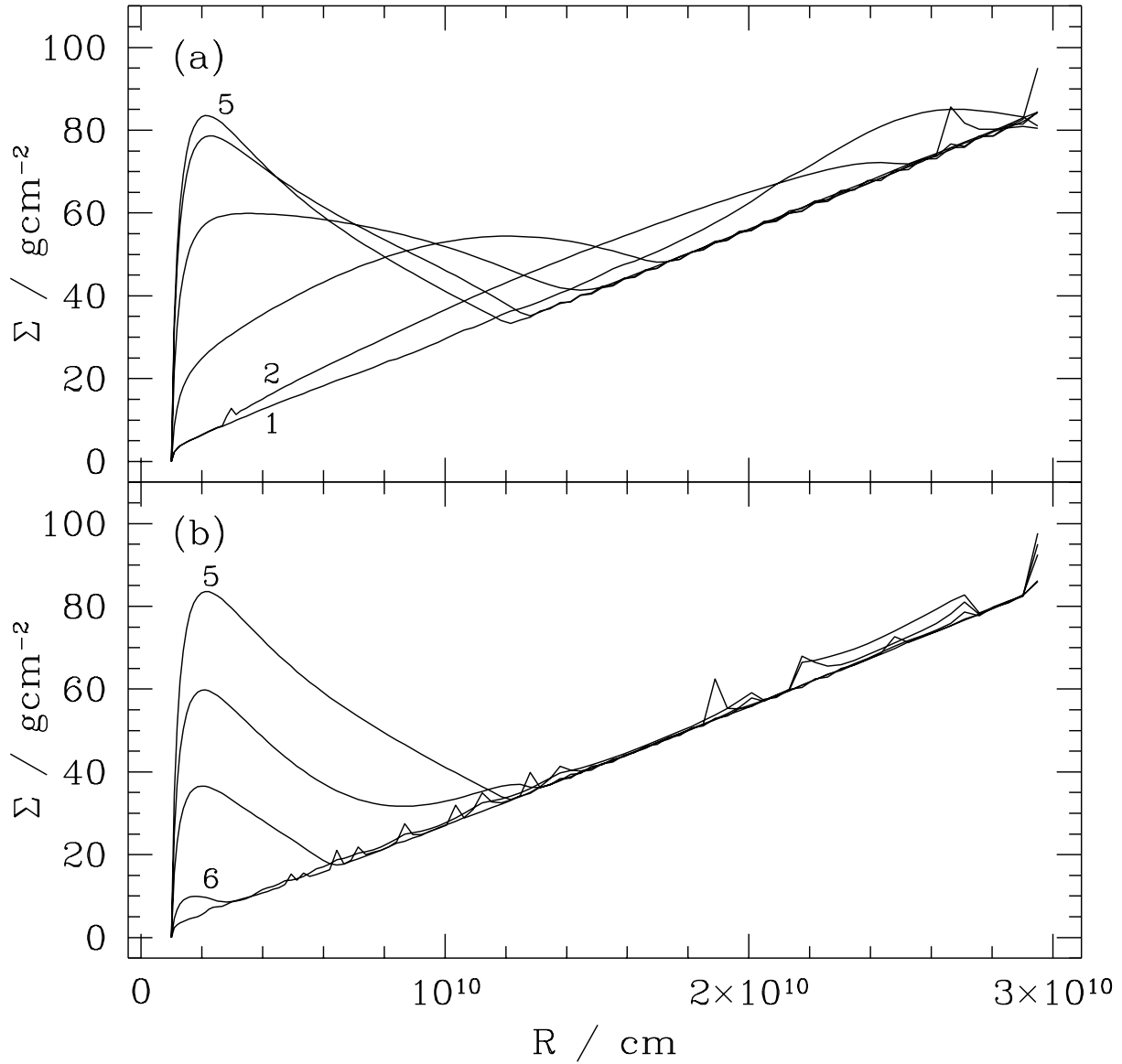

Fig. 4.— The evolution of the disc surface density profile. (a) On the rise to outburst and (b) on the decline. Curves show ($\Sigma/\mathrm{gcm^{-2}}$) snapshots taken at $2 \times 10^4$s intervals on the rise, and at $10^5$s intervals on the decline. The numeric labels identify curves according to the position of the inner disc zones in the log($\Sigma/\mathrm{gcm^{-2}}$) - log($T_e/\mathrm{K}$) plane (Fig. 3).

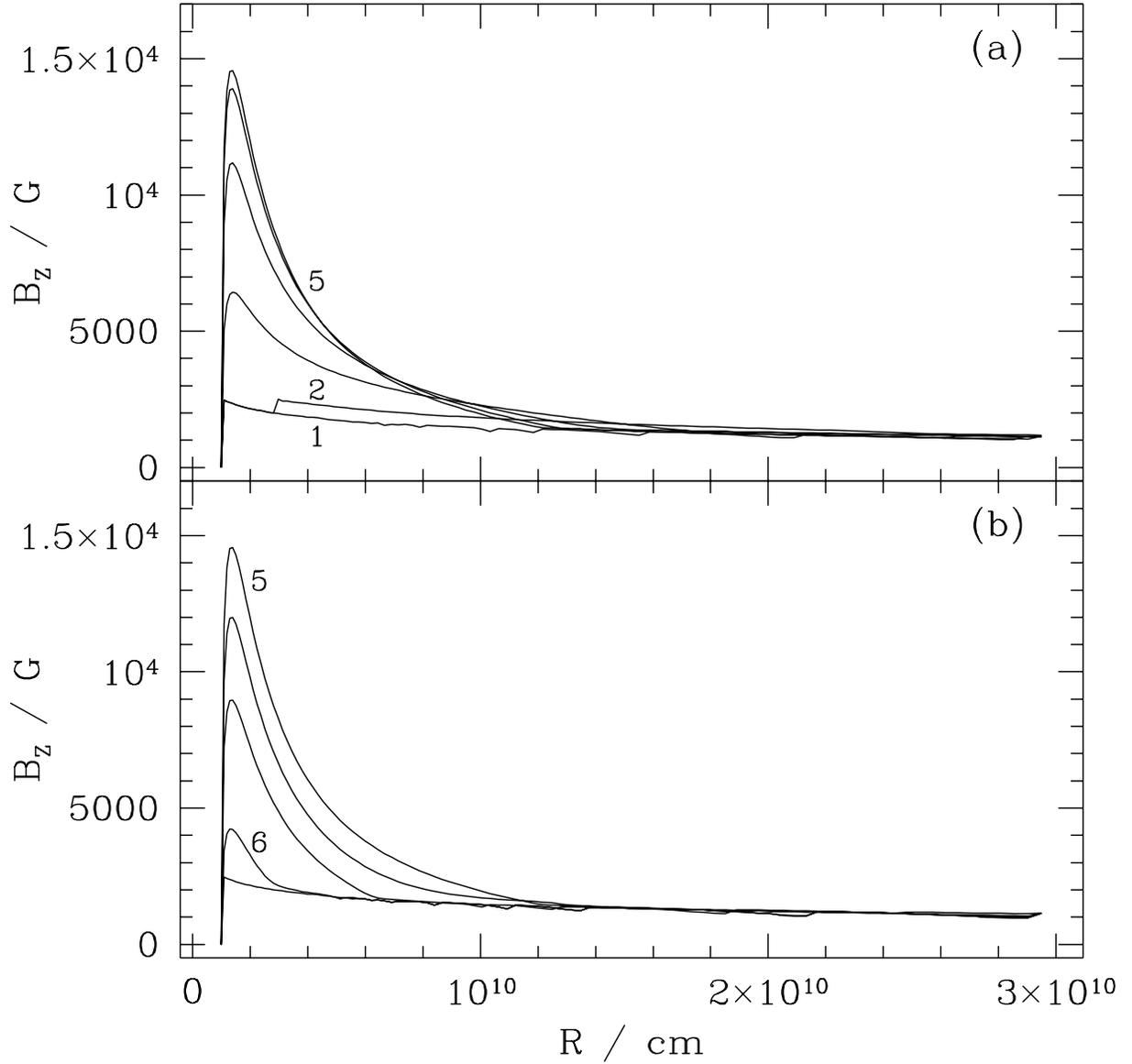

Fig. 5.— The evolution of the vertical component of the magnetic field. (a) On the rise to outburst and (b) on the decline. Curves show $(B_Z/G)$ snapshots taken at $2 \times 10^4$s intervals on the rise, and at $10^5$s intervals on the decline. The numeric labels identify curves according to the position of the inner disc zones in the $\log(\Sigma/\mathrm{gcm^{-2}})$ - $\log(T_e/\mathrm{K})$ plane (Fig. 3).

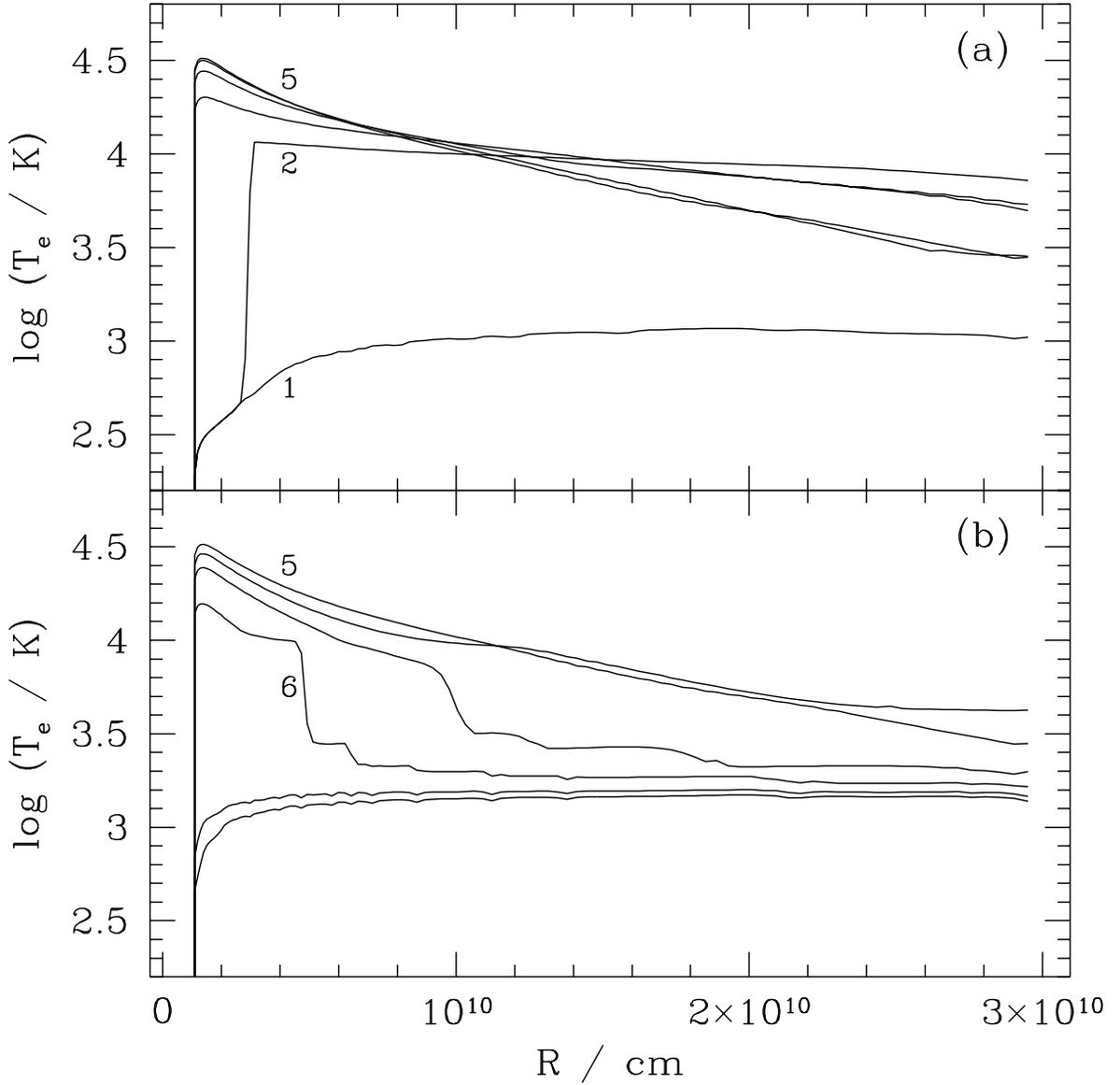

Fig. 6.— The evolution of the disc effective temperature. (a) On the rise to outburst and (b) on the decline. Curves show $\log(T_e/K)$ snapshots taken at $2 \times 10^4$s intervals on the rise, and at $10^5$s intervals on the decline. The numeric labels identify curves according to the position of the inner disc zones in the $\log(\Sigma/\mathrm{gcm^{-2}})$ - $\log(T_e/K)$ plane (Fig. 3).